\documentclass[reprint, twocolumn,notitlepage, superscriptaddress, nofootinbib]{revtex4-2}

\usepackage[english]{babel} 
\usepackage{microtype} 
\usepackage[svgnames]{xcolor} 
\usepackage{graphicx} 
\usepackage{enumitem} 
\setlist{noitemsep} 
\usepackage{lastpage}
\usepackage{fancyhdr}
\usepackage{lipsum}

\usepackage{geometry} 
\usepackage[T1]{fontenc} 
\usepackage[utf8]{inputenc} 
\usepackage{XCharter} 
\newcommand{\rev}[1]{\textcolor{black}{#1}}

\renewcommand{\Re}[0]{\text{Re}}
\renewcommand{\Im}[0]{\text{Im}}
\newcommand{\7}[1]{\pmb{#1}}

\usepackage{natbib}
\usepackage[nottoc]{tocbibind}
\addto\captionsenglish{}
\usepackage[autostyle=true]{csquotes} 

\usepackage{amsmath}
\usepackage{amsfonts}
\usepackage{amssymb}
\usepackage{textgreek}
\usepackage{cleveref}
\usepackage{braket}
\usepackage{bbold}
\usepackage{dsfont}
\usepackage{amsthm}

\usepackage{listings}
\usepackage[compat=1.0.0]{tikz-feynman}
\usepackage{feynmf}
\usepackage{tikz}
\usetikzlibrary{decorations.markings}
\tikzset{
	photon/.style={decorate, decoration={snake}, draw=black},
	particle/.style={thin,draw=black, postaction={decorate},
		decoration={markings,mark=at position .5 with {\arrow[black]{stealth}}}},
	density/.style={decorate, draw=black, decoration={snake=coil}},
	propagator/.style={line width=0.75mm, draw=black, postaction={decorate}, decoration={markings,mark=at position .5 with {\arrow[black]{stealth}}}},
}
\tikzset{
	photon/.style={decorate, decoration={snake}, draw=black},
	particle/.style={thin,draw=black, postaction={decorate},
		decoration={markings,mark=at position .5 with {\arrow[black]{stealth}}}},
	density/.style={decorate, draw=black, decoration={snake=coil}},
	propagator/.style={line width=0.75mm, draw=black, postaction={decorate}, decoration={markings,mark=at position .5 with {\arrow[black]{stealth}}}},
}
\usepackage{fancyhdr} 
\usepackage{lastpage}
\begin{document}
	\title{Vibrational phenomena in glasses at low temperatures captured by
		field theory of disordered harmonic oscillators}
	
	\author{Florian Vogel}
	\author{Matthias Fuchs} \affiliation{University of Konstanz - D-78457 Konstanz, Germany} 
	
	\begin{abstract}
		\textbf{Abstract-} We investigate the vibrational properties of topologically disordered materials by analytically studying particles that  harmonically oscillate around random positions. Exploiting \rev{classical} field theory  in the thermodynamic limit at $T=0$, we build up a self-consistent model by analyzing the Hessian utilizing Euclidean Random Matrix theory. In accordance with earlier findings 
  [T. S. Grigera et al.J.~Stat.~Mech.~
11 (2011) P02015.],
we take non-planar diagrams into account to correctly address multiple local scattering events. By doing so, we end up with a first principles theory that can predict the main anomalies of athermal disordered materials, including the boson peak, sound softening, and  Rayleigh damping of sound. 
		In the vibrational density of states, the sound modes lead to Debye's law for small frequencies.
  Additionally, an excess appears in the density of states starting as $\omega^4$
in the low frequency limit, which is attributed to (quasi-) localized modes.
	\end{abstract}

	\date{\today} 

	\begin{widetext}
		
		\maketitle 
		
	\end{widetext}

	\textit{Introduction.--}
	The athermal excitations in glasses differ characteristically from the ones in ordered systems of the same chemical substances. While the vibrational properties of crystalline solids are well understood in terms of phonons,  
 \rev{viz.~wave-like small} particle displacements from lattice positions, the vibrational spectra of amorphous solids exhibit incompletely understood anomalies.

	One usually names three phenomena \cite{SchirmacherVibrationalAnoamlies}. I) Whereas the Debye law holds in crystalline solids in the low energy regime, there appears a maximum in the reduced vibrational density of states (vDOS) $\frac{g(\omega)}{\omega^2}$ in amorphous solids \cite{ciliberti2003brillouin, SchirmacherVibrationalAnoamlies, PhysRevE.52.4026, PhysRevLett.104.195501, Franz2015}. This maximum is referred to as the \textit{boson peak}, where   $\omega$ is the frequency. II) Experimental and computational data suggest that the sound attenuation results from disorder-scattering and  is Rayleigh-like $\propto \7p^4$ below the boson peak, where $\7p$ is the wave vector. When entering the frequency regime of the boson peak the damping turns  into a $\7p^2$-law \cite{Monaco3659Breakdown, Monaco16907,Schrimacher2007,MarruzzoSchirmacher, Ikeda_Phonon_transport, PhysRevLett.104.195501, PhysRevLett.112.125502, BaldiEmergence} which is additionally indicated by a III) softening of the sound velocity, i.e.~a dip in the reduced dispersion relation around the frequency of the boson peak \cite{Monaco16907, Monaco3659Breakdown, Schirmacher_Heterogeneous_Elasticity}. It has been conjectured that these phenomena are interrelated and that they are connected to quasi-localised modes (QLMs) \cite{Schrimacher2007, Schirmacher_Heterogeneous_Elasticity, Ikeda_Phonon_transport, Wang_Stable_glasses, Szamel2022, Mahajan2021,Lerner_2021}. QLMs have been found in many computer simulations of disordered materials. It was also demonstrated that their density of states follows a universal $\propto \omega^4$ law and that they hybridize with phonons, so that neither of the two modes are exact eigenvectors of the dynamical matrix anymore, which \rev{is constituted by} the Hessian of the potential energy \cite{Ikeda_Phonon_transport, Lerner_2021, NonpnononicSpectrum}. 
	
	The localisation of modes in amorphous systems and the resulting fluctuations of elastic constants is at the heart of many prominent models, such as the two-level system \cite{Anderson1972}, the soft potential model \cite{Karpov_LocalisedModes, Parshin_1993, Parshin_2007} and its generalizations \cite{Bouchbinder}, mean field approaches \cite{Franz2015,DeGiuli2014}, and the heterogeneous elasticity theory (HET) \cite{Schrimacher2007, Schirmacher_Heterogeneous_Elasticity, Mahajan2021}. Nevertheless, all these approaches require phenomenological parameters and they generally do not capture the vibrational anomalies starting from the microscopic laws of motion. For example, the widely used HET \cite{Schrimacher2007, Schirmacher_Heterogeneous_Elasticity, Mahajan2021} is a mesoscopic rather than a microscopic theory which quantitively underestimates the importance of QLMs \cite{Wang_Stable_glasses, Szamel2022, Caroli2019}.

	In this work, we start from the microscopic equations of disordered coupled harmonic oscillators. This approach leads to the euclidean random matrix (ERM) problem suggested by Parisi and co-workers \cite{M_zard_1999,Martin_Mayor_2001, ciliberti2003brillouin}. Following them, we rely on a Green's function formalism, to derive a self-consistent model that rationalises all aforementioned anomalies and thus improves on earlier ERM-models. The guiding principle in our derivation is that multiple local scattering events are of qualitative importance \cite{sheng1995introduction}. This is also hinted at by the discovered influence of non-planar diagrams  \cite{Leutheusser1983, grigera2011high}, \rev{which  were identified as origin of Rayleigh damping in the ERM \cite{Ganter_Schirmacher,Ganter2011,Schirmacher2019}.} Therefore, we develop a model that relies on a vertex instead of propagator renormalization. 
	\\

	\textit{The system.--} 
We study a system of $N$ particles randomly placed in a  volume $V$ at the positions $\{\7r_i\}^N$ in the thermodynamic limit with  $N/V$ being constant. The positions are drawn from a uniform distribution $P[\{\7r_i\}^N]=1/V^N$. Considering small fluctuations $\phi_i$ around the 
frozen positions $\7r_i$, we define the symmetric random matrix $\7M$ via the second derivative of an interaction pair potential
$U(\{\phi_i\}) = \frac{1}{2} \sum_{i,j=1} f(\7r_i - \7r_j ) (\phi_i - \phi_j)^2= \sum_{i,j} M_{ij} \phi_i \phi_j\;.$
The $f$ is a spring function which quantifies the interaction strength. We only request for the theoretical investigation that the Fourier transformation  $\hat{f}(\7p)$ exists.
We also assume rotational invariace, so that $\hat{f}$ only depends  on the absolute modulus of the wavevector $p= |\7p|$ and that the  spring function is regular.  This implies $\hat{f}(\70)-\hat{f}(\7p) \propto \7p^2$ for small $\7p$. When performing numerical calculations, we set $\hat{f}(\7p)= (2 \pi \sigma^2)^{3/2} e^{-\sigma^2 p^2/2}$. Here $\sigma$ is an intrinsic length scale of the system, which leads to a dimensionless density $n=N\sigma^3/V$. The density $n$ turns out to be the single state parameter. In the following, $\sigma$ will be set to unity.  Note, that we neglect the vector character of $\phi$. The scalar $\phi$s represent transverse displacements\rev{, which predominantly contribute to the boson peak \cite{Horbach2001}}.\\

The fundamental equations of motion of $N$ coupled harmonic oscillators read
\begin{align}
	\label{Equation_of_motion}
	\ddot{\phi}_i=- \sum_{j=1}^N\; M_{ij}\;  \phi_j\;, \quad\mbox{for } 1\le i \le N\;.
\end{align} 
Here, time and (later) frequency are made dimensionless by a frequency scale $\omega_0$ \rev{(set to $\omega_0=1$ for simplicity)} that can be taken from the position of the boson peak in measurements.
Translational invariance and hence momentum conservation follow immediately from the potential $U(\{\phi_i\})$.  Consequently, $\7M$ has the eigenvalue zero. The associated eigenvector $\7e_0$ corresponds to the uniform shift  $\7e_0=(1,1,....,1)$. \rev{For positive spring function, the potential $U$ is positive and thus  the matrix $\7M$ is semi-positive definite. }

It is noteworthy, that the disorder in $\7M$ and the thermodynamic limit lead to a broadening of the oscillator lines in the dynamic structure factor and to sound attenuation, even though the eigenvalues of \rev{the matrix} $\7M$  are exclusively \rev{non-negative and thus the oscillator frequencies real}. We interpret this as a instantiation of Landau damping \cite{Mouhot2011}: \rev{In  time-reversible equations of motion and in the thermodynamic limit, damping can arise from energy transfer among the infinite multitude of modes.} 

We study the ERM system by analysing the two-point response or Green's function $G$. 
\rev{It gives the evolution of an initial displacement field with plane wave form of wavevector $\7p$. $G(\7p,z)$ is its spectrum at eigenvalue $z$ and} is related to the   resolvent of $\7M$
\begin{align}
	\label{Def_Resolvent}
	\begin{split}
		G(\7p,z)&=  \underset{N,V \to \infty}{\text{lim}} \frac{1}{V} \overline{\sum_{i,j=1}^N e^{i \7p\cdot(\7r_i-\7r_j)} \left[\frac{1}{z-\7M}\right]_{ij}} \;.
	\end{split}
\end{align}
Here $z =(\omega+i0^+)^2\in \mathbf{C}$  with  $\omega$ corresponding  to the frequency. The overline indicates the sample average over the disorder. 
The resolvent can be connected to observables  like the dynamic structure factor  and the density of states \cite{goetschy2013euclidean, ciliberti2003brillouin, M_zard_1999}. See the supplemental material (SM), Sect.~\ref{App_Observables} \cite{SI},
for further information. \\

\textit{Self-consistent model.--} \label{Sec_Self_consistent_Model}
Following \cite{Martin_Mayor_2001, ciliberti2003brillouin, grigera2011high}, we perform a high density expansion of the resolvent \eqref{Def_Resolvent}. Using the Dyson equation, $G=G_0+  G_0\, \Sigma\, G$, 
we express the Green's function in terms of a bare propagator $G_0(\7p,z)=[z/n-\epsilon_0(\7p)]^{-1}$ and the self energy $\Sigma(\7p,z)$, with $\epsilon_0(\7p)=\hat{f}(\70)-\hat{f}(\7p)$ giving the bare dispersion relation. \rev{ While $G_0$ describes undamped harmonic oscillators, $\Sigma$ arises from the disorder in the elastic couplings. We envision a perturbation traveling through the system, and consider the field $\phi_i$  as excitation at the respective lattice site so that the interaction between the perturbation and the disorder can be called scattering event \cite{grigera2011high}. The self energy thus contains }
all the inelastic scattering events.  $\Sigma$ has a series expansion in $\frac{1}{n}$ and vanishes for $n \to \infty$, where the disorder vanishes. \rev{Thus, $1/n$ quantifies the disorder and the weakening of the elastic constants $f(\7 r_i-\7r_j)$ when the separation of particles gets larger. } Using Feynman diagrams, we reconstruct the different inelastic scattering processes.  Since this approach has been tried before \cite{ciliberti2003brillouin,grigera2011high, Martin_Mayor_2001}, we moved further comments on the technical details to the SI, Sect.~\ref{Sub_Feynman}. 

The derivation of our self-consistent model starts  with the  insight, that any contribution to the self energy necessarily ends with the same vertex and  that  the momentum is conserved at every vertex. This allows us to write down the self energy schematically:

\begin{align}\label{vertex}
	&\Sigma(\7p,z)= 
	\vcenter{\hbox{\begin{tikzpicture}
				\draw[particle] (0,0) -- (0.89,0);
				\fill[](0,0) circle (0.1);
				\node at (1,0) [rectangle,draw] {};
				\draw[photon] (0,0) arc(180:12.3:0.51) ;
	\end{tikzpicture}}}\; \\& 
	\vcenter{\hbox{\begin{tikzpicture}
				\draw[photon] (0.03,0.51) arc(90:11.9:0.5);
				\draw[particle] (0,0) -- (0.39,0);
				\node at (0.5,0) [rectangle,draw] {};
	\end{tikzpicture}}}=
	\vcenter{\hbox{\begin{tikzpicture}
				\draw[photon] (0.0,0.5) arc(90:11.5:0.5);
				\draw[particle] (0,0) -- (0.42,0);
				\fill[](0.5,0) circle (0.1);
	\end{tikzpicture}}}+ \underset{A}{\vcenter{\hbox{\begin{tikzpicture};
					\draw[particle] (0 ,0) -- (0.45,0);
                    \draw[particle] (0.45,0) -- (1,0);
					\draw[particle] (1.03,0) -- (1.37,0);
					\fill[](0.38,0) circle (0.1);
					\fill[](0.95,0) circle (0.1);
					\draw[photon] (0.38,0) arc(180:0:0.294) ; 
					\node at (1.48,0) [rectangle,draw] {};
					\draw[photon] (0.977,0.538) arc(90:11.9:0.54) ;
	\end{tikzpicture}}}}+
	\underset{B}{\vcenter{\hbox{\begin{tikzpicture}
					\draw[particle] (0 ,0) -- (0.45,0);
                    \draw[particle] (0.45,0) -- (1,0);
					\fill[](0.87,0) circle (0.1);
					\fill[](0.4,0) circle (0.1);
					\draw[photon] (0.45,0.035) arc(180:10.0:0.46) ;
					\draw[photon] (0.15,0.53) arc(110:0:0.5) ;
					\node at (1.37,0) [rectangle,draw] {};
					\draw[particle] (1,0) -- (1.25,0) ;
	\end{tikzpicture}}}} + 
	\underset{C}{\vcenter{\hbox{\begin{tikzpicture}
					\draw[particle] (0 ,0) -- (0.45,0);
                     \fill[](0.45,-0.0) circle (0.1);
					\draw[particle] (0.50 ,0) -- (1.21,0);
					\draw[photon] (0.5,0.026) arc(180:12.7:0.432) ;
					\draw[photon] (-0.0,0.35) arc(90:0:0.4) ;
					\node at (1.32,0) [rectangle,draw] {};
	\end{tikzpicture}}}}+ \cdot \cdot \cdot			\notag
\end{align}												
    Here, a straight line represents the bare propagator;  a curly line  a density fluctuation and the circle denotes a vertex and marks an inelastic scattering event. The square can be regarded as a renormalized vertex \cite{Chattopadhyay2000}, which absorbs all possible  insertions at a bare vertex.  The letters $A,$ $B$, $C$ just label the different building blocks in \eqref{vertex} which are of second order in density fluctuation. The three dots represent more diagrams with more simultaneous density fluctuations. Every new loop comes with an additional factor $1/n$. Thus, one can truncate the expansion after a few orders in the high density limit.
																									
    A self-consistent model is easily  constructed by only keeping  classes of diagrams with the  topologies  $A$, $B$, and $C$ and, in the lower line of Eq.~\eqref{vertex}, by replacing the bare propagator between two bare  vertices  with the full Green's function $G(\7p,z)$. 
(Note, a \rev{dressed} Green's function ending in a renormalized vertex would lead to an overcounting. This can be easily seen by inserting the Dyson equation.)
		In contrast to earlier models \cite{ciliberti2003brillouin,Grigera2001,Schirmacher2019},  this re-summation takes all the diagrams that topologically match the ones from second order perturbation theory and hence  non-planar diagrams into account.  We do this for two related reasons: I) QLMs are arguably important for the modes of vibrations of low temperature glasses \cite{Lerner_2021, Wang_Stable_glasses, Szamel2022, Schober_Localised_modes} and one must therefore correctly consider multiple local scattering events. Planar diagrams  underestimate these scattering sequences \cite{sheng1995introduction, Wang_Stable_glasses, Szamel2022}. II) Non-planar diagrams are needed to give the correct  Rayleigh-damping of sound modes for $p \to 0$ and to prevent a potential infrared divergence  of the self energy \cite{Leutheusser1983, Schnyder_2011}. 
			The Feynman rules stated in the SI, Sect.~\ref{Sub_Feynman}, allow to write down the associated amplitude for our self-consistent model 
\begin{subequations}
		\label{Self_consitent_model}
	\begin{align}
				G&(\7p,z)= \frac{1}{z/n- \epsilon_0(\7p)- \Sigma(\7p,z)}\;, \\
		\Sigma&(\7p,z) = \int \frac{d^3 \7k}{(2 \pi)^3} V(\7k,\7p) \mathcal{V}(\7k,\7p,z)\;, \\
\notag	\Big[ n & G_0^{-1}(\7k,z)-\int \frac{d^3\7q}{(2 \pi)^3}V^2(\7q,\7k) G(\7q,z) \Big] \mathcal{V}(\7k,\7p,z) \\
\notag &= V(\7k,\7p) +
	\int \frac{d^3\7q}{(2 \pi)^3}\Big[ V(\7q-\7k,2 \7q-\7p)+V(\7p-\7q,\7k)\\&\times G(|\7p-\7k-\7q|,z)V(\7p-\7k,\7q) \Big] \mathcal{V}(\7q,\7p,z) \;,
\label{EquationXi} \\
\label{SymmetryVertex}
V&(\7k,\7p)=\hat{f}(\7k)-\hat{f}(\7k-\7p)=-V(\7p-\7k,\7p)\; .
					\end{align}
			\end{subequations}
While one could easily include more diagrams, there is no need for it. On the contrary, we will now argue that this minimal model successfully captures all the vibrational phenomena of low temperature glasses. \rev{Importantly, we consider stable glass states while previous approaches had considered marginally stable glasses where a close-by instability leads to vibrational anomalies \cite{ciliberti2003brillouin,Shimada2020,Grigera2001}.}\\
										
			\textit{Results.-- a) Dispersion relation:} 
		The dispersion relation $\epsilon(\7p)=n(\epsilon_0(\7p)+\text{Re}[\Sigma(\7p,z=0]))$ characterizes the peak positions of the vibrational modes in the dynamic structure factor; it is shown in Fig.~\ref{DispersionFM}. \rev{The limiting proportionality $\epsilon(p)\propto n$ for $n\to\infty$ arises from the pairwise interaction among all particles.}
			The expansion \rev{$\epsilon(\7p \to 0) \to (c_T\,p)^2$ } indicates the presence of sound waves  in the hydrodynamic limit. They are expected as Goldstone modes arising from broken translational invariance. \rev{Here,  $c_T$ is the    (transverse) speed of sound.} 
			Lowering the density increases the disorder and weakens the elasticity; the frequencies of vibrations become softer.
  \rev{Additionally, a} dip \rev{appears} around $\sigma p \approx 10$.  This indicates a negative dispersion of the sound velocity, \textit{i.e} sound softening, and also suggests the presence of the boson peak in the vDOS \cite{Monaco16907, Monaco3659Breakdown, Schirmacher_Heterogeneous_Elasticity}. \rev{For very small $n$, $\epsilon(p\approx10/\sigma)$ may become negative, but this density range is not considered.} 
   \rev{Considering only diagrams of type $A$ in Eq.~\eqref{vertex}, a  self-consistent re-summation of all planar diagrams is possible \cite{ciliberti2003brillouin,Grigera2001}, which for reference is presented in the SI, Sect.~\ref{app_sec_Planar_theory}. It captures wave modes equally well and gives comparable results for $\epsilon(p)$ as included in Fig.~\ref{DispersionFM}.}
			\begin{figure}
		\centering
\includegraphics[scale=1]{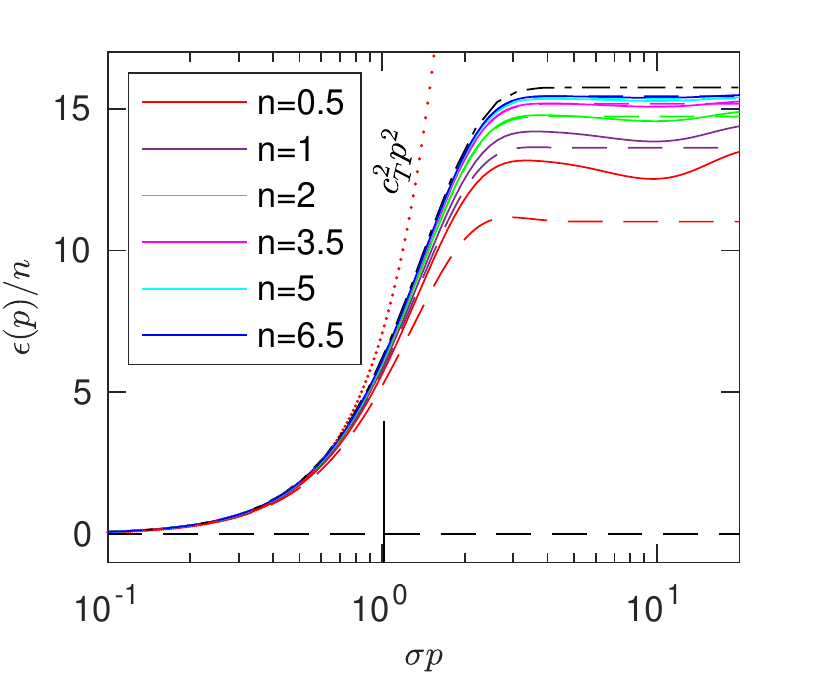}
		\caption{
            The reduced dispersion relation $\epsilon(p)/n$ (solid lines) is shown for different densities as function of wavevector $p$. It is compared to the associated bare dispersion $\epsilon_0(p)$ (dashed-dotted line) \rev{and to the result from the planar re-summation (dashed lines) \cite{Grigera2001}.} \rev{The limit of sound propagation, $\epsilon(p\to0)\to(c_Tp)^2$ is indicated for $n=0.5$.}
           \rev{At this $n$, the vertical bar marks  $p_{BP}= \omega_{BP}/c_T$, the wavenumber delimiting the sound behavior. }}
				\label{DispersionFM}
	\end{figure}

		\textit{b) Sound attenuation:}
		The sound attenuation  is given by the imaginary part of the self-energy. It determines the width of the vibrational mode around the sound pole. 
		The self-consistent re-summation of the planar diagrams alone \cite{Grigera2001,ciliberti2003brillouin,grigera2011high} leads to strong hydrodynamic sound damping (viz.~$\propto p^2$), while experiments \cite{BaldiEmergence,Monaco3659Breakdown} and simulations \cite{ContinuumLimitMizun,Monaco16907,Wang_Stable_glasses}  indicate weaker Rayleigh damping  (viz.~$\propto p^4$). It can be understood to result from wave scattering off the frozen disorder.
		To show that the non-planar diagrams fix the error of a planar self-consistent approach, we argue that the imaginary parts of the planar diagrams (class $A$, first line in Eq.~\eqref{EquationXi} and given in  diagram (\ref{SoundAttenuationVertices}a)) and non-planar diagrams (class $B$, last line in Eq.~\eqref{EquationXi} and given in diagram (\ref{SoundAttenuationVertices}b)) cancel each other exactly for $\7p \to 0$. 
\begin{subequations}
\label{SoundAttenuationVertices}
\begin{align}
A= \vcenter{\hbox{\begin{tikzpicture}
\node[] (A) at (0,0) {} ;
\node[] (B) at (1.0,0) {} ;
\node[] (C) at (3,0) {};
\node[] (D) at (4,0) {};
\draw[particle] (A) -- node[below]{$\7p$} (B);
\draw[propagator] (B) -- node[above]{$\7q$} (2,0);
\draw[propagator] (2,0) -- node[below]{$\7k$} (D);
\draw[particle] (3.95,0) -- node[below]{$\7q$} (5,0);
\draw[photon] (0.95,0.05) -- node[right] {$\7p-\7q$} (2,-1);
\fill[](B) circle (.12);
\fill[](2,0) circle (.12);
\fill[](D) circle (.12);
\draw[photon] (2,0) arc(180:0:1) node at (3,1.5) {$\7q-\7k$} (B);
\end{tikzpicture}}} \;,
\end{align}
\begin{align}
\hspace{-0.425cm} B= \vcenter{\hbox{\begin{tikzpicture}
\node[] (A) at (0,0) {} ;
\node[] (B) at (1.0,0) {} ;
\node[] (C) at (3,0) {};
\node[] (D) at (4,0) {};
\node[] (E) at (3.8,0.8) {};
\draw[particle] (A) -- node[below]{$\7p$} (B);
\draw[propagator] (B) -- node[below]{$\7p+\7k-\7q$} (C);
\draw[propagator] (2.9,-0.1) -- node[above]{$\7k$} (3.9,0.8);
\draw[particle] (E) -- node[above]{$\7q$} (5,0.8);
\draw[photon] (2.95,-0.05) -- node[right] {$\7p-\7q$} (4.2,-1.2);
\fill[](B) circle (.12);
\fill[](C) circle (.12);
\fill[](3.8,0.75) circle (.12);
\draw[photon] (B) arc(180:30:1.5) node at (2.3,1.95) {$\7q-\7k$} (B);
\end{tikzpicture}}}\;.
\end{align}
\end{subequations}
The thick line represents the full Green's function. Both diagrams describe equivalent scattering processes  off two density fluctuations  but in different sequence.
The cancellation can be seen by applying the Sokhotski-Plemelj identity $[x\pm i0^\pm]^{-1}=\mp i \pi \delta(x)+\mathcal{P}\left(\frac{1}{x}\right)$ to the full propagator in the hydrodynamic limit and by integrating over $\7k$;  here, $\mathcal{P}$ represents the Cauchy-principal value. For small $\7p$, the  symmetry  \eqref{SymmetryVertex} gives the cancellation; see proof in the SI, Sect.~\ref{Supp_Prof_Rayleigh}. 
	It also fixes the infrared divergence problem  \cite{Leutheusser1983, Schnyder_2011}. 
	The building block containing the four vertex (diagram \textit{C} in (S5) in the SI)  gives the correct imaginary part by itself.
	In total, this leads to   $ G(\7p,z)/n=[z-\epsilon(\7p)-i\rev{\omega(\7p)}\Gamma(\7p)]^{-1}$ with $\Gamma(\7p)=n \Im\Sigma(\7p,z\!=\!\epsilon(\7p)) / \rev{\omega(\7p)}=B_{\rm R}\, p^4$ around the sound pole $\omega(\7p)=\sqrt{\epsilon(\7p)}$ in the hydrodynamic limit. The strength of Rayleigh damping $B_{\rm R}$ increases with disorder.

	\begin{figure}
	\centering
\includegraphics[scale=1]{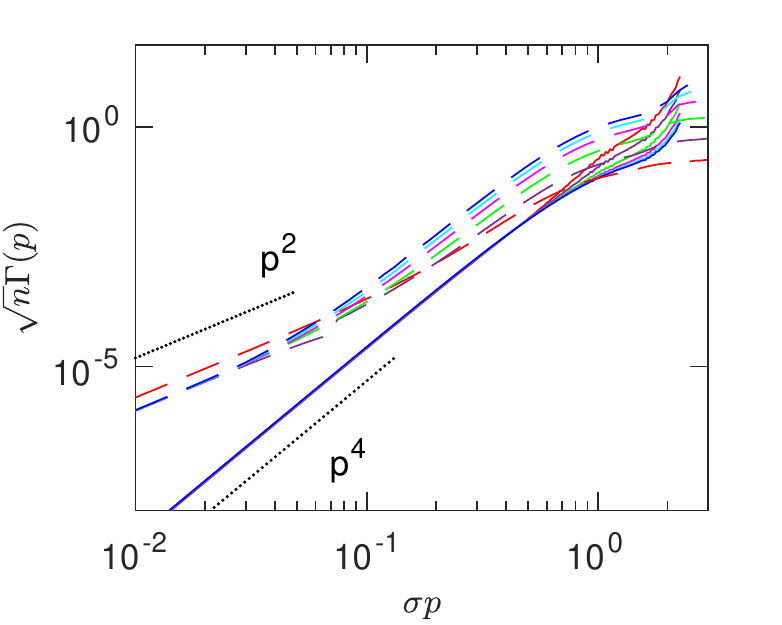}
	\caption{ Sound attenuation  as function of wavevector $p$. \rev{Rescaled data $\sqrt{n} \Gamma(p)$ collapse for high densities $n$ (see legend in Fig.~\ref{DispersionFM}) for small $p$.} \rev{Solid lines follow from} the imaginary part of the self-energy \rev{given by Eq.~\eqref{Self_consitent_model}, dashed lines follow from planar diagrams \cite{Grigera2001} (see SI).} The sound attenuation is calculated around the sound pole $\omega=\sqrt{\rev{n} \epsilon_0(p)}$. \rev{Dotted lines represent asymptotic power laws. }}
			\label{fig:Damping}
	\end{figure} 
													
	Figure  \ref{fig:Damping} shows the sound attenuation for different densities in the two loop approximation; see the SI for details.
	Since our full model \eqref{Self_consitent_model} topologically coincides with the second order, the second order solution confirms, that \eqref{Self_consitent_model} predicts the correct sound attenuation.

	\textit{c) Vibrational density of states:}
	The  vDOS can be calculated from the large wavevector limit of the Green's function where only diagonal elements of $\7 M$ contribute in Eq.~\eqref{Def_Resolvent} \cite{Martin_Mayor_2001,ciliberti2003brillouin}; see  SI, Sect.~\ref{Supp_vDOS}, for details.
	The sound modes already identified in the dispersion relation suggest  that the vDOS contains a Debye spectrum    
 $g_D(\omega) =  \omega^2 / \omega_D^3$  for   $\omega \to 0$. 
    \rev{The Debye frequency $\omega_D$ characterizes the region of long-wavelength sound and gives  an upper cut-off for waves in solids}. It shrinks with increasing disorder and the magnitude of the Debye law increases for decreasing $n$; see panel a) in Fig.~\ref{fig:vDOS}. Note, that panel $a)$ has been calculated under the assumption that $\omega^2$ is small; see SI, Sect.~\ref{Supp_vDOS}. 
    This approximation breaks down for $\omega \to 1$. The boson peak is situated at the upper end of the spectrum of  vibrations in the model. There, the vDOS can be simplified as the contributions of the acoustic phonons to the self energy become weak. This leads to  a closed expression for the vDOS which is  Wigner's semi circle law as expected in uncorrelated random matrix ensembles  \cite{Gotze2000,goetschy2013euclidean,Franz2015}.   The \rev{amplitude of the} boson peak  shown in panel c) of Fig.~\ref{fig:vDOS} \rev{only varies little} with increasing disorder, while its position shifts trivially with $\sqrt n$.   \rev{The ratio of its position to the Debye-frequency, $\omega_{BP}/\omega_D$ (see the inset in panel c) of Fig.~\ref{fig:vDOS}), is smaller as one indicating that $\omega_{BP}$, and not $\omega_D$, sets the limit for wave behavior in random matrix approaches \cite{Schirmacher_Heterogeneous_Elasticity,Conyuh2021}. 
    In simulations of stable glasses  \cite{2019NatureCommunication}, the boson peak lies low, $\omega_{BP}/\omega_D\approx 0.17$.} 
	\begin{figure}
	\centering
		\includegraphics[scale=0.8]{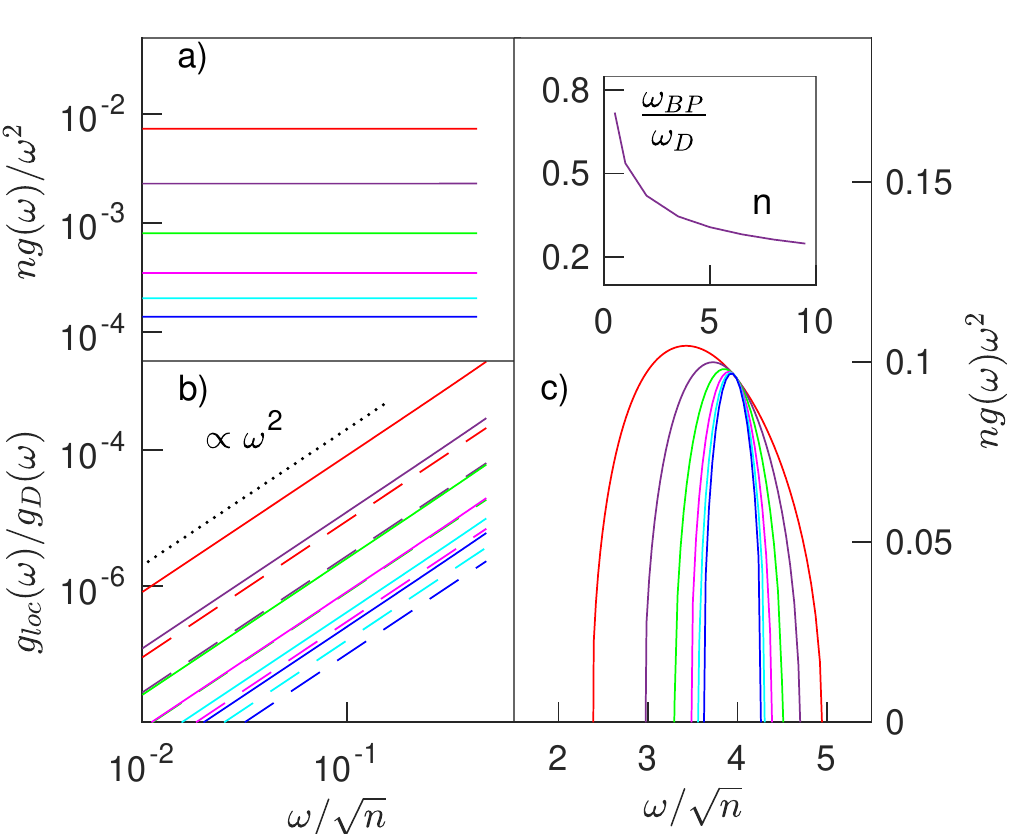}
	\caption{ Panel a), full lines show the reduced vibrational density of states (vDOS), {$\rev{n}g(\omega)/\omega^2$,}  for low frequencies at different number densities $n$. 
 Panel b) presents the vDOS of the quasi-localised modes (QLM), \rev{}{$g_{\rm loc}(\omega)/g_D(\omega)$,} where the dashed line shows the prediction of the HET theory. Panel c) exhibits the \rev{rescaled} boson peak, \rev{$n g(\omega)/\omega^2$}, which is located at the upper end of the dispersion relation. \rev{The inset shows the ratio $\omega_{BP}/\omega_D$ of boson peak and Debye frequencies.}
\rev{ The densities and their respective colours are the same  in all three panels following the legend in Fig.~\ref{DispersionFM}}.}
\label{fig:vDOS}
\end{figure} 	
			
\textit{d) Quasi localised modes:}
Recent works  \cite{Wang_Stable_glasses, 2019NatureCommunication, Schrimacher2007, Lerner_2021}  established a close relation between QLMs and  Rayleigh-damping by showing that there is a linear relation between the damping coefficient  $B_R$  and the coefficient $A_4$ of the characteristic  vDOS of the quasi localised modes $g_{\rm loc}=A_4\omega^4$. Additionally, it was argued in \cite{SCHOBER2011} that the  presence of QLMs implies a $p^4$ sound attenuation. Furthermore, it has been shown that QLMs give  rise to the boson peak \cite{Ikeda_Phonon_transport, Wang_Stable_glasses, Lerner_2021, SchirmacherVibrationalAnoamlies, Parshin_2007}. This suggests that QLMs are at the heart of the vibrational anomalies of disordered materials.  Our results in Figs.~\ref{fig:Damping} and \ref{fig:vDOS} support this narrative. 
	In finite systems, the participation ratio can be used to identify QLMs, which is impossible here as the thermodynamic limit was taken.  Thus, we interpret the QLMs as the modes that have a vDOS  proportional  to the Rayleigh term $B_R$.  We show the quartic contribution to the vDOS in panel b) of Fig.~\ref{fig:vDOS}, again utilising  a small $\omega$ approximation. We also compare it to the HET-prediction $g^{\rm HET}_{\rm loc}/\omega^4= \rev{2 B_R/(  \pi \omega_D^2c_T^4)}$ \cite{Schrimacher2007,Ikeda_Phonon_transport}, which underestimates  disorder \rev{in stable glasses} quantitatively \cite{Wang_Stable_glasses,2019NatureCommunication}, \rev{where $(c_T^4 \omega_{BP}^2) A_4 / B_R \approx 0.05$ holds; our ratio   $0.045$ for $n=0.5$ lies close.}
 \rev{The anomaly is missing in the vDOS of the self-consistent planar theory \cite{ciliberti2003brillouin,Grigera2001}, which confirms that planar diagrams overly restrict the sequence of interactions of vibrational modes with particle sites; for details see the SI, Sect.~\ref{app_sec_Planar_theory}. }\\
													
		\textit{Conclusion and Outlook.--}
	Our self-consistent field theory of ERM accounts for disorder
	more accurately than approaches based on mean field or coherent potential approximations. The latter underestimate multiple local scattering events, which become important if one has bound states or localisation  effects  \cite{sheng1995introduction}. Neglecting dependent scattering processes in an ERM model leads to a planar theory for the vDOS in the thermodynamic limit \cite{goetschy2013euclidean}.   This, together with the cancellation of diagrams in Eq.~\eqref{SoundAttenuationVertices}  to get the correct Rayleigh damping suggests  that non-planar diagrams are essential to correctly address disorder.
	Besides this qualitative insight, we constructed a self consistent theory for disordered harmonic oscillators that correctly predicts all the vibrational anomalies of disordered materials. \rev{It can be coarse-grained and then leads to the widely-used HET.} After understanding the topology of athermal disorder, the next step is to take the vector character of the displacement fields into account and to consider finite temperatures.
	Additionally, it would be worthwhile to relate the approach to the soft potential model and its generalizations.

	\textit{Acknowledgment.--} We thank Matthias Krüger for the co-supervision of the Master's thesis that was the basis for this work,  and  him, Annette Zippelius, Thomas Franosch\rev{, and Walter Schirmacher} for fruitful discussions.   The work was supported  by the Deutsche Forschungsgemeinschaft (DFG) via SFB 1432 project CO7.

\medskip
\bibliography{sources}
\clearpage
\title{Supplemental Materials for "Vibrational phenomena in glasses at low temperatures captured by field theory of disordered harmonic oscillators"}
\maketitle
\onecolumngrid
\setcounter{equation}{0}
\setcounter{figure}{0}
\setcounter{table}{0}
\setcounter{section}{0}
\setcounter{page}{1}
\makeatletter
\renewcommand{\theequation}{S\arabic{equation}}
\renewcommand{\thefigure}{S\arabic{figure}}

\section{Feynman rules \label{Sub_Feynman}}
The main focus in this paper lies on the averaged Green's function, which is the Fourier and Laplace transformed  two-point response  function with  $-s^2=z$, where $s=-i \omega + 0^+$ is the Laplace frequency 
\begin{align}
G(\7p,z) = \rev{- \frac{1}{V}}\frac{1}{s} \sum_{i,j=1}^N\int d^3 \7r_i d^3 \7r_j P_2(\7r_i,\7r_j) \int_0^{\infty}e^{-i\7q\cdot(\7r_j-\7r_i)}e^{-st} \braket{\phi(\7r_i,t),\phi(\7r_j,0)}\;,
\end{align}
with $\Re[z]>0$ and $\phi(\7r_i,t)$ being the elongation of the i$^{th}$ particle from its equilibrium position. $\braket{\cdot \cdot \cdot}$ representing the average  in a given quenched disorder. Since we consider zero temperature, the average in the quenched disorder can be pictured as sending a sound mode through the same configuration of particles with different starting positions  and then taking the average of the different two-point functions. The integrals over the positions $\{\7r_i\}^N$ give the average over the disorder. Here,  $P_2(\7r_i,\7r_j)$ is the joint probability distribution of the particle locations. In this paper, we assume uniform and independent distributions $P_2(\7r_i,\7r_j)=P_1(\7r_i)P_1(\7r_j)=1/V^2$. \\
													
Unfortunately, one can only calculate the two-point function analytically in exceptional cases. Hence, one relates the interacting system to the non-interacting system using the Gell-Mann Low theorem \cite{Coleman}.
Assuming that the interaction is weak and turned on adiabatically, one can describe the full system as a perturbation of the non-interactive system. The resolvent of the latter, the bare propagator is denoted by $G_0$. The interaction can be interpreted as inelastic scattering events of a vibrational wave with the disorder that causes   density fluctuations. The associated Feynman rules have been lucidly derived in \cite{grigera2011high} using a field theoretical and a combinatorial approach. Thus,  we just state the Feynman rules and refer to the cited paper for the derivation.  \\

\begin{minipage}[b]{0.2\textwidth}
\raggedright
\begin{itemize}
\item Field Excitation:
\vspace{0.6cm}
\item Density Fluctuation:
\vspace{0.8cm}
\item Three Vertex:
\vspace{1.4cm}
\item Four Vertex:
\vspace{0cm}
\end{itemize}
\end{minipage}
\begin{minipage}[b]{0.3\textwidth}
\centering
$\hspace{0.4cm}\vcenter{\hbox{\begin{tikzpicture} 
[node distance=1 cm and 1.5 cm]
\coordinate[] (e1);
\coordinate[ right=of e1] (aux1);
\draw[particle] (e1) -- node[above]{\7p} 
(aux1); 
\end{tikzpicture}}} $ \\
\vspace{0.5cm}
$\hspace{0.4cm}\vcenter{\hbox{\begin{tikzpicture} 
[node distance=1 cm and 1.5 cm]
\coordinate[] (e1);
\coordinate[ right=of e1] (aux1);
\draw[photon] (e1) -- node[above]{\7q} 
(aux1); 
\end{tikzpicture}}}$ \\
\vspace{0.5cm}
$\hspace{0.31cm}\vcenter{\hbox{\begin{tikzpicture}
[node distance=1 cm and 2 cm]
\node[] (C) at (0.5,0) {};
\node[] (D) at (-0.0,0.6) {};
\node[] (A1) at (1.5,1) {};
\draw (D) [photon] arc(90:0:0.7) node[xshift=-0.5cm, yshift=1cm]{$\7p-\7q$} (C);
\draw (-0,0.0) [particle] -- node[below]{$\7q$}(0.75,0);
\draw (0.75,0) [particle] --node[below]{$\7p$}(1.5,0);
\fill[](0.75,0) circle (.13);
\end{tikzpicture}}}$
$\hspace{0.645cm}\vcenter{\hbox{\begin{tikzpicture}
[node distance=1 cm and 2 cm]
\node[] (C) at (0.5,0) {};
\node[] (D) at (-0.5,1) {};
\node[] (A1) at (1.5,1) {};
\draw (D) [photon] arc(90:0:1) node[xshift=-1.0cm, yshift=1.3cm]{$\7p-\7q$} (C);
\draw (C) [photon] arc(180:90:1) node[xshift=0.cm, yshift=0.3cm]{$\7p-\7k$} 
(A1); \draw (-0.5,0.0) [particle] -- 
node[below]{$\7q$}(0.5,0);
\draw (C) [particle] --
node[below]{$\7k$}(1.5,0);
\fill[](0.5,0) circle (.13);
\end{tikzpicture}}}$
\end{minipage}
\begin{minipage}[b]{0.4\textwidth}
\raggedright
\begin{subequations}
\begin{align}
\hspace{-2.cm}
=G_0(\7p,z)= \frac{1}{z\rev{/n}-\epsilon_0(\7p)}
\end{align} \\
\vspace{-0.2cm}
\begin{align}
\hspace{-2.35cm}=1\hspace{2.5cm}
\end{align}
\vspace{0.02cm}
\begin{align}
\hspace{-0.3cm} =V(\7q,\7p)= \hat{f}(\7q)-\hat{f}(\7q-\7p) \end{align} 
\vspace{0.2cm}
\begin{align}
\hspace{-2.6cm} =V(\7q-\7k,2\7q-\7p)
\end{align}
\vspace{-0.8cm}
\end{subequations}
\end{minipage} \\
Note , that we use a slightly different notation compared to \cite{ciliberti2003brillouin, grigera2011high}. We pulled out the density to make the dependence on this parameter more transparent.
While the bare propagator only considers   stiff connections between the particles,   the full Green's function also takes  scattering processes into account. The difference between the two propagators is the self-energy $\Sigma(\7p,z)=G_0^{-1}(\7p,z)-G^{-1}(\7p,z)$, which hence encodes the full complexity of the system. 
\section{Observables \label{App_Observables}}
The two observables considered in this work are the dynamic structure factor $S_\lambda$ and the vibrational density of states $g_\lambda$. Both can be related to the Green's function. $S_\lambda$ measures the correlation between the position of the i$^{th}$ particle at time $0$ and the j$^{th}$ particle at time $t$. The subscripted $\lambda$ refers to the  eigenvalue space, in which the dynamic structure factor reads
\begin{align}
\label{def_dynamic_strucutre_factore}
\begin{split}
S_\lambda(\7p, \lambda) &= \underset{N\to \infty}{\text{lim}} \frac{1}{2N} \sum_n \sum_{i,j=1}^N \overline{e^{i \7p\cdot( \7r_j- \7r_i)} e_n(i)e_n(j) \delta(\lambda-\lambda_n) } \\& =- \frac{1}{n \pi} \Im G(\7p,\lambda+i 0^+) \;.
\end{split}
\end{align}
Here, the overline denotes the average for an arbitrary but fixed disorder, $\lambda=\omega^2=z$ represents the energy values and $\lambda_n$ are the eigenvalues of $\7M$. \rev{The Hessian matrix is determined by the spring function, $M_{ij} = \delta_{ij} \sum_k f(\7r_i-\7 r_k)-f(\7r_i-\7r_j)$.}  The dynamic structure factor exhibits peaks around the sound poles $z=\rev{n}(\epsilon_0(\7p)+\Re\Sigma(\7p,z))$ with the self energy $\Sigma$ introduced in Eq.~(\ref{Self_consitent_model}a).  The width of these peaks is given by the imaginary part of the self energy, which hence determines the sound attenuation. \\
													
The density of states is given by the trace of the resolvent and \rev{thus} by the high momentum limit of the dynamic structure factor \cite{ciliberti2003brillouin, grigera2011high, Martin_Mayor_2001}.
This gives the expression
\begin{align}
\begin{split}
g_\lambda(\lambda)\equiv \frac{1}{N} \sum_n^N \delta(\lambda-\lambda_n)=- \frac{1}{n \pi} \underset{p \to \infty}{\text{lim}} \Im G(p,\lambda+i 0^+).
\end{split}
\end{align}
The density of states in the frequency domain is obtained by multiplying $g_\lambda(\lambda)$ with the Jacobian $\left| \frac{\partial \lambda}{\partial \omega} \right|$. This gives $ g(\omega) = 2\omega g_\lambda(\lambda(\omega))\;.$ \\
													
\section{Rayleigh Damping of the self-consistent model \label{Supp_Prof_Rayleigh} }
It has  already been proven in \cite{grigera2011high} that the Euclidean random matrix model without any approximations predicts Rayleigh-damping. In this section, we prove  that  our self consistent model \eqref{Self_consitent_model} does so as well. The idea is that the imaginary part of the self-energy in the lowest order arises from the imaginary part of exactly one of the propagators constituting a diagram times the real part of the remaining ones. \\

Considering an arbitrary diagram it is self-evident from \eqref{Self_consitent_model} that every diagram from second order onwards can be built up from the following three building blocks plus   an initial and a final vertex $V(\7q',\7p)$ or respectively $V(\7k',\7p)$
													
\begin{align}
\label{SoundAttenuationVertices_appen}
A= \vcenter{\hbox{\begin{tikzpicture}
\node[] (A) at (0,0) {} ;
\node[] (B) at (1.0,0) {} ;
\node[] (C) at (3,0) {};
\node[] (D) at (4,0) {};
\draw[particle] (A) -- node[below]{$\7p$} (B);
\draw[propagator] (B) -- node[above]{$\7q$} (2,0);
\draw[propagator] (2,0) -- node[below]{$\7k$} (D);
\draw[particle] (3.95,0) -- node[below]{$\7q$} (5,0);
\draw[photon] (0.95,0.05) -- node[right] {$\7p-\7q$} (2,-1);
\fill[](B) circle (.12);
\fill[](2,0) circle (.12);
\fill[](D) circle (.12);
\draw[photon] (2,0) arc(180:0:1) node at (3,1.5) {$\7q-\7k$} (B);
\end{tikzpicture}}} 
\hspace{0.5cm} B= \vcenter{\hbox{\begin{tikzpicture}
\node[] (A) at (0,0) {} ;
\node[] (B) at (1.0,0) {} ;
\node[] (C) at (3,0) {};
\node[] (D) at (4,0) {};
\node[] (E) at (3.8,0.8) {};
\draw[particle] (A) -- node[below]{$\7p$} (B);
\draw[propagator] (B) -- node[below]{$\7p+\7k-\7q$} (C);
\draw[propagator] (2.9,-0.1) -- node[above]{$\7k$} (3.9,0.8);
\draw[particle] (E) -- node[above]{$\7q$} (5,0.8);
\draw[photon] (2.95,-0.05) -- node[right] {$\7p-\7k$} (4.2,-1.2);
\fill[](B) circle (.12);
\fill[](C) circle (.12);
\fill[](3.8,0.75) circle (.12);
\draw[photon] (B) arc(180:30:1.5) node at (2.3,1.95) {$\7q-\7k$} (B);
\end{tikzpicture}}}
\hspace{0.5cm}C= \vcenter{\hbox{\begin{tikzpicture}
\node[] (A) at (0,0) {} ;
\node[] (B) at (1.0,0) {} ;
\node[] (C) at (3,0) {};
\node[] (D) at (4,0) {};
\draw[particle] (A) -- node[below]{$\7p$} (B);
\draw[propagator] (B) -- node[below]{$\7k$} (C);
\draw[particle] (2.95,0.05) -- node[below]{$\7q$} (4,1);
\draw[photon] (2.95,-0.05) -- node[right] {$\7p-\7q$} (4,-1);
\fill (B) circle (.12);
\fill (C) circle (.12);
\draw[photon] (B) arc(180:0:1) node at (2,1.5) {$\7p-\7k$} (B);
\end{tikzpicture}}}\;.
\end{align}
The crucial point is, that the imaginary part of the $A$- and $B$-blocks cancel each other out exactly.   As a consequence, the lowest surviving order of the imaginary part of the self energy is Rayleigh-like $\Im \Sigma \propto  \sqrt{\lambda}p^4$. To make the building-block structure appear in the equation for the renormalised vertex, we  pull out a factor $G_0^{-1}(\7k,z) $ from the renormalised vertex  $\mathcal{V}(\7k,\7p,z) \to G_0^{-1}(\7k,z) \mathcal{V}(\7k,\7p,z)$.  The self-consistent equations for the self energy hence read 
\begin{align}
\label{Xi2}
\begin{split}
\Sigma(\7p,z)=\sum_{\7k} V(\7k&,\7p) G_0(\7k,z) \mathcal{V}(\7k,\7p,z),\\
\mathcal{V}(\7k,\7p,z)= \frac{V(\7k,\7p)}{n}&+ \frac{1}{n} \sum_{\7q} \bigg\{ \sum_{\7l} V^2(\7l,\7k)G(\7l,z)\delta(\7k-\7q)  \\&+ V(\7q-\7k,2\7q-\7p)+V(\7p-\7q,\7k)G(\7p-\7q-\7k,z)V(\7p-\7k,\7q)\bigg\}G_0(\rev{\7q},z) \mathcal{V}(\7q,\7p,z)\;,
\end{split}
\end{align}
with $\sum_{\7 k}=\int \frac{d\7k}{(2\pi)^3}$.
For simplicity of notation, integrals over wavevectors are abbreviated as summations throughout the supplemental information.
We  will now analyse this equation term by term by applying the Sokhotski-Plemelj formula  \begin{align}
\label{Sokhotski-Plemelj formula }
\frac{1}{x+i0^+}=P\frac{1}{x}- i \pi \delta(x)\;,
\end{align} 
with $P$ denoting the Cauchy Principal value.		\begin{itemize}
\item[1.)] The first term leads to the first order perturbation theory
\begin{align}
\begin{split}
\frac{1}{n} \sum_{\7k} V(\7k,\7p)&G_0(\7k,\lambda+i0^+)\mathcal{V}(\7k,\7p,\lambda+i0)^{(1)} = - \frac{1}{n} \sum_{\7k} V(\7k,\7p)G_0(\7k,\lambda+i0^+)V(\7k,\7p)
\end{split}
\end{align}
Applying \eqref{Sokhotski-Plemelj formula } gives the correct order of eigenvalue and momentum in the hydrodynamic limit, \textit{i.e.} $\Im\Sigma^{(1)}= a \lambda^\frac{1}{2}p^4 + b \lambda^\frac{3}{2}p^2$, with $a,b \in \mathbb{R}$. The notation $\mathcal{V}^{(m)}$ denotes the $m^{th}$ term in equation \eqref{Xi2}.
\item[2.)] The next term is more complicated since the renormalised vertex $\mathcal{V}$ appears left and right of the equality sign. But again, the imaginary part results from a single loop in the hydrodynamic limit. Everything else would lead to higher orders in $\lambda$ since the imaginary part of any loop vanishes with the eigenvalue. Hence, one can look at a single term/loop in $\mathcal{V}$. By doing so, one leaves it open what might appear to the left or right of it. Applying \eqref{Sokhotski-Plemelj formula } yields
\begin{align}
\Im\mathcal{V}(\7k,\7p,\lambda+i0^+)^{(2)}=& \frac{1}{n} \sum_{\7q} V(\7q-\7k,2 \7q-\7p) \Im G_0(\7q,\lambda+i0^*) \mathcal{V}(\7q,\7p,\lambda)\\&\underset{\lambda\to 0}{\longrightarrow} -\frac{ \pi S_3}{2 (2\pi)^3c_0^2 } \left(\frac{\lambda}{c_0^2}\right)^\frac{1}{2}V(\7k,\7p)\mathcal{V}(0,\7p,\lambda) \propto \lambda^\frac{1}{2}\7p \mathcal{V}(0,\7p,\lambda)\;,
\end{align}
with the speed of sound $c_0= \underset{q \to 0}{\text{lim}} \frac{\sqrt{n\epsilon_0(\7q)}}{q}$ of the bare system and  $S_3$ being the surface  area  of the unit sphere.
Since the initial and final two vertices both give an additional factor $\7p$, the contribution to the imaginary part of the self-energy arising from the imaginary part of $\mathcal{V}^{(2)}$ is at least of order $\lambda^\frac{1}{2}\7p^3$. But such a term can not appear due to momentum inversion invariance. So the leading order must be at least $\lambda^\frac{1}{2}\7p^4$. \\

Note that one could also apply \eqref{Sokhotski-Plemelj formula } to the bare propagator $G_0(\7k,z)$ in
\begin{align}
\frac{1}{n }\sum_{\7k,\7q} V(\7k,\7p) G_0(\7k,z)V(\7q-\7k,2 \7q-\7p) G_0(\7q,z) \mathcal{V}(\7q,\7p,\lambda)\;.
\end{align} Here, one gets a similar result, as one can easily check.
\item[3.+4.)] The contributions from the $A$- and $B$-diagrams are of the wrong order, but the two terms cancel themselves in the hydrodynamic limit which gives the correct overall sound attenuation. Again, one has to focus on a single loop and disregard everything to the left. Note, that applying \eqref{Sokhotski-Plemelj formula } to the bare propagators in \eqref{EquationXi} gives no \rev{significant} contribution due to the identity $V(\7q,0)=0$. So, the imaginary part arises from the terms containing the dressed propagator. For the resolvent one can use
\begin{align}
\Im G(\7q,\lambda+i0^+) = \frac{\Im\Sigma(\7q,\lambda)}{\Big(\lambda/\rev{n}-\epsilon_0(\7q)-\Re\Sigma(\7q,\lambda)\Big)^2+ \Im\Sigma(\7q,\lambda+i0^+)^2} \underset{\lambda \to 0}{\longrightarrow}- \rev{ \pi \frac{n}{2c_T^2q} \delta( \sqrt{\lambda/c_T^2 }}-  q) \;.
\end{align}
After performing the integral over $\7q,\7l$ one can calculate the contribution from the third and fourth term of the vertex $\mathcal{V}$ for vanishing $\7p$. One gets
\begin{align}
\begin{split}
\Im \mathcal{V}^{(3)}(\7k,\7p,\lambda+i0^+) =& \frac{1}{n}\sum_{\7l}V(\7l,\7k)^2\Im\Big(G(\7l,\lambda+i0^+)\Big)G_0(\7k,\lambda)\mathcal{V}(\7k,\7p,\lambda) \\&\underset{\lambda\to 0 }{\longrightarrow} \frac{ \pi S_3}{\rev{2}(2\pi)^3  c_T^2} \left(\frac{\lambda}{c_T^2}\right)^\frac{1}{2} \epsilon_0(\7k) \mathcal{V}(\7k,\7p,\lambda)\;.
\end{split}
\end{align}
One can use a similar line of argumentation to investigate the imaginary part of $\mathcal{V}^{(4)}.$ The contribution in the small $\7p$ limit reads
\begin{align}
\begin{split}
\Im \mathcal{V}^{(4)} (\7k,\7p,\lambda +i0^+) =& \frac{1}{n} \sum_{\7q} V(-\7q,\7k)\Im\Big(G(\7q+\7k),\lambda+i0^+)\Big)V(-\7k,\7q)G_0(\7q,\lambda)\mathcal{V}(\7q,\7p,\lambda) \\&\underset{\lambda\to 0 }{\longrightarrow} \frac{ \pi S_3}{\rev{2}(2\pi)^3  c_T^2} \left(\frac{\lambda}{c_T^2}\right)^\frac{1}{2}\epsilon_0(\7k) \mathcal{V}(-\7k,\7p,\lambda)\;.
\end{split}
\end{align}
Adding both terms together yields
\begin{align}
\label{ImContributionAB}
\Im\Big(\mathcal{V}^{(3)}(\7k,\7p,\lambda)+\mathcal{V}^{(4)}(\7k,\7p,\lambda)\Big) \underset{\7p,\lambda\to 0}{\longrightarrow}  \frac{ \pi S_3}{\rev{2}(2\pi)^3  c_T^2} \left(\frac{\lambda}{c_T^2}\right)^\frac{1}{2} \frac{\epsilon_0(\7k)}{n} \bigg(\mathcal{V} (\7k,\7p,\lambda )+\mathcal{V} (-\7k,\7p,\lambda ) \bigg)\;.
\end{align}
The important point is that the bracket vanishes in the hydrodynamic limit. This follows from the definition $V(\7q,\7p)= \hat{f}(\7q)-\hat{f}(\7q-\7p)$ and $\hat{f}(\7p)=\hat{f}(-\7p)$ which holds due to rotational invariance . This gives
\begin{align}
\begin{split}
V(-\7k,\7p\to 0 ) &= \7p \cdot \frac{\partial \hat{f}(\7k)}{\partial (- \7k)}= - V(\7k,\7p\to 0 )\;, \\
V(\7q,\7k)&= V(-\7q,-\7k,).
\end{split}
\end{align}
So, ignoring all initial momenta in the second, third and fourth term in $\mathcal{V}$ and changing the integration variable $\7q \to- \7q$ gives in the small $\7p$-limit for the vertex $\mathcal{V}$ \eqref{EquationXi}
\begin{align}\begin{split}
\mathcal{V}(-\7k,\7p,z)&= -\frac{V(\7k,\7p)}{n} + \frac{1}{n} \sum_{\7q} \bigg( V(\7q-\7k,2 \7q) +\sum_{\7l}V(\7l,\7q) G(\7l,z) V(\7l,\7q)\delta(\7k-\7q) \\&\hspace{0.2cm} + V(-\7k,\7q) G(\7k+\7q,z)V(-\7q,\7k) \bigg) G_0(\7q,z) \mathcal{V}(-\7q,\7p,z) =- \mathcal{V}(\7k,\7p,z).
\end{split}
\end{align}
This holds since the only change the minus sign in the argument on the left-hand side inflicts on the right-hand sight is in the final vertex $V(\7k,\7p)$. But this vertex marks the end of any considered diagram. Hence, it appears exactly once in every occurring term. Thus, the imaginary contributions from $\mathcal{V}^{(3)}$ and $\mathcal{V}^{(4)}$ in \eqref{ImContributionAB} cancel each other in the lowest order in $\7p$. Consequently, the leading order must be $\lambda^{\frac{1}{2}}\7p^4$ or $\lambda^\frac{3}{2}\7p^2$\rev{, respectively}. \\
\end{itemize} 
In the main paper, we numerically calculate the imaginary part of the self energy around the sound pole $\lambda=\epsilon(\7p)$ in the two loop approximation. That is, truncating the equation for the renormalised vertex \eqref{EquationXi} by replacing $\mathcal{V}$ on the right hand side with the non-renormalised vertex $V$ and setting $G\equiv G_0$ in the $A$- and $B$- blocks. The resulting diagrams are the ones from the second order perturbation theory. Since the used building blocks  \eqref{SoundAttenuationVertices_appen} topologically coincides with the blocks of the second order perturbation theory, this numerical solution already implies that our full model predicts the correct sound attenuation. The analytical proof confirms this formally. 
													
\section{Analytical Result for the vibrational Density of states \label{Supp_vDOS}}
In this section, we analytically investigate the vDOS predicted by our model. The density of states in the eigenvalue domain is given by 
\begin{align}
g_\lambda(\lambda)=\frac{1}{N}\sum_{n=1}^N \rev{\delta(\lambda-\lambda_n)}= -\frac{1}{N\pi} \Im \overline{ \sum_{i=1}^N \left[\frac{1}{\lambda+i0^+ - \7M} \right]_{ii}}\;.
\end{align}
The trace of the resolvent is thus related to the high momentum limit of the \rev{propagator}.  The  self-consistent equation for the high momentum limit  read
\begin{align}
\begin{split}
G_\infty( z)&\equiv G(\infty,z)=[z\rev{/n}-\hat{f}(0)-\Sigma(\infty,z)]^{-1}\\
\Sigma(\infty,z)&= \sum_{\7k} \hat{f}(\7k)\Big[\mathcal{V}(\7k,\infty,z)- \mathcal{V}'(\7k,\infty,z)\Big]\;,
\end{split}
\end{align}
with $\mathcal{V}'(\7k,\infty,z)= \mathcal{V}(\7k\to\7p +\7k,\7p=\infty,z)$. 
The   two renormalized vertices obey
\begin{align}
\label{Term1}
\begin{split}
\bigg(n G_0^{-1}(\7k,z) - \underbrace{\sum_{\7q} V^2(\7q,\7k)G(\7q,z)}_{A} \bigg) 	\mathcal{V}(\7k,\infty,z)=  \hat{f}(\7k)+ \sum_{\7q} \bigg\{	\underbrace{\hat{f}(\7q-\7k)}_{C}\mathcal{V}(\7q,\infty,z)\\- \Big[ \underbrace{ \hat{f}(\7q+\7k)}_{C} + \underbrace{V(-\7q,\7k) G(\7k+\7q,z)\hat{f}(\7k+\7q)}_{B}   \Big]\mathcal{V}'(\7q,\infty,z)
\end{split}
\end{align}
and
\begin{align}
\label{Term2}
\begin{split}
\bigg(n G_0^{-1}(\infty,z) &-  \underbrace{ \sum_{\7q}  \hat{f}(\7q)^2 \Big[ G(\7q,z) +G_{\infty}(z)}_A \Big]  \bigg) 
\mathcal{V}'(\7k,\infty,z)=-\hat{f}(\7k) \rev{-} \sum_{\7q} \bigg\{ \Big[ \underbrace{  \hat{f}(\7q+\7k)}_C \\&+\underbrace{ \hat{f}(\7k+\7q)  G(\7q+\7k,z)V(-\7k,\7q) }_B \Big] \mathcal{V}(\7q,\infty,z) + \Big[ 
\underbrace{\hat{f}(\7q-\7k)}_C+  \underbrace{\hat{f}(\7q)G_{\infty}(z)\hat{f}(\7k)}_B  \Big] \mathcal{V}'(\7q,\infty,z)  \bigg\} \;.
\end{split}
\end{align}
Here, the letters \textit{A,B,C} represent the respective building block \eqref{SoundAttenuationVertices_appen} from which the  term originates.
													
	\paragraph{High frequency limit and semi-circle law}
	The boson-peak occurs at the upper limit  of the dressed dispersion relation. For large eigenvalues, one has $|G(\7k,z)| \approx |\rev{n}/z| \ll 1$ for $\epsilon(\7k) \ll \epsilon(\infty)$. Thus, we ignore all contributions to the self energy $\Sigma(\infty,z) $ that depend on $G(\7k,z)$ or $G_0(\7k,z)$. Looking at the equations for the renormalised vertex \eqref{Term2}, one notices, that the \textit{A} and \textit{B}  blocks contribute with the same coefficient $a=\frac{1}{n}\sum_{\7k} \hat{f}(\7k)^2 $ in the considered approximation, if one ignores the $C$-diagrams. \rev{ For the considered Gaussian spring function holds $a = \frac{\hat{f}(\70)}{\sqrt{8}n}$}. This gives the following  self consistent equation for the high frequency limit of the resolvent 
\begin{align}
G_\infty^{-1}(z)= G_0^{-1}(\infty,z)- \frac{1}{2} \sum_{m=1}^\infty (2 a)^m G_0(\infty,z)^m G_\infty(z)^{m-1}-\Sigma_C(z)\;,
\end{align}
where $\Sigma_C$ represents  all the terms that contain at least one \text{C}-block. The term $2^{m-1}$ is a combinatorial factor.  Since the geometric sum converges for $n$ being sufficiently large, we get
\begin{align}
\label{Semi-Circle_equation}
1= \rev{2}G_0^{-1}(\infty,z)G_\infty(z)- \frac{1}{1-2aG_\infty(z)G_0(\infty,z)}- \rev{2}\Sigma_C(z)G_\infty(z)
\end{align}
For large $n$, we can approximate $\Sigma_C$ by the first two elements of perturbation theory $\Sigma_C=\Sigma_C^1+\Sigma_C^2+\cdot \cdot \cdot.$.
\begin{align}
\Sigma^1_C(z)&=\frac{G_0(\infty,z)^2}{n^2} \sum_{\7k,\7q}\hat{f}(\7k)\hat{f}(\7k-\7q)\hat{f}(\7q) \\
\Sigma^2_C(z)&=\frac{G_0(\infty,z)^3}{n^3} \sum_{\7k,\7q,\7l}\hat{f}(\7k)\hat{f}(\7k-\7q)\hat{f}(\7q-\7l)\hat{f}(\7l)\;.
\end{align}
Since the dressed dispersion relation is smaller than the bare one, we set $G_0(\infty,z) \approx - 1/\hat{f}(\70)$. Note, that this assumption is essential for the convergence of the infinite sum. This gives the quadratic equation
\begin{align}
1+\bigg(\frac{ a}{\hat{f}(\70)}-G_0^{-1}(\infty,z)+\Sigma_C(\infty,z)\bigg)G_\infty(z)+2aG_\infty(z)^2\approx 0\;,
\end{align}
\rev{where we neglected higher terms in $\Sigma_C G(\infty)$.}
It has the solution 
\begin{align}
G_\infty(z)=\frac{G_0^{-1}(\infty,z)-b \pm \sqrt{(z\rev{/n}-\hat{f}(0)-b)^2-8a}}{4a}
\end{align}
with $b=\Sigma_C+\frac{a}{\hat{f}(0)}$. This gives the semi-circle law for the distribution of eigenvalues centered around $\omega_{BP}^2/n= \hat{f}(0)+b$ and with a width $\sqrt{8a}$. \rev{The maximal height  of the resulting peak  in the vDOS  is hence $g_{\omega, max}=\frac{\sqrt{2} \omega_{BP}}{\pi n \sqrt{a}} \propto n^{0}$. That the height of the peak only depends on the bare spring function is an artefact of the applied approximations. Additionally,} the ignored terms lead to a shift of this distribution to smaller frequencies. \rev{ For $n$ sufficiently high, one can approximate the dispersion relation with the bare dispersion. This gives for the relation of maximal height of the peak and the Debye level 
$\frac{g_{\omega,max}}{\omega_D^3} \propto n^{-\frac{5}{2}} $ for sufficiently high densities.} \\
													
\paragraph{Low frequency limit and Debye law} 
Contrary to the high frequency limit, for $z \to 0$, we can approximate $| G(\infty,z)| \approx 1/\hat{f}(0) \ll 1$. Setting terms equal to zero that contain at least one factor $G_0(\infty,z)$ highly simplifies the equations for the renormalised vertex. The amplitude of the remaining diagrams is given by 
\begin{align}
\Sigma(\infty,z) &= \sum_{\7k} \hat{f}(\7k) \mathcal{V}(\7k,\infty,z)\;,\\
\mathcal{V}(\7k,\infty,z) &= \frac{\hat{f}(\7k)+\sum_{\7q} \hat{f}(\7k-\7q)  \mathcal{V}(\7q,\infty,z)}{nG_0^{-1}(\7k,z)-\sum_{\7q} V(\7q,\7k)^2 G(\7q,z)}\;.
		\end{align}
	Note, that the $B$-blocks are only implicitly present via the dressed propagator $G(\7q,z)$.
	Since high momentum contributions  of $G(\7q,z)$ to the denominator are exponentially suppressed, one can approximate $G$ with the hydrodynamic limit of the resolvent $\rev{n}G^{-1} \approx z-\epsilon(\7q)+i \omega(q)^3 q^2 \rev{\frac{B_{\rm R}}{c_T^2}}. $ Here, we also added the two leading contributions to the damping together  $ \omega^3 q^2  \rev{B_{\rm R}/c_T^2} \approx   \omega^3 q^2 B_1 + \omega q^4 B_2$ which holds close to the sound pole $z \approx \epsilon(\7q)$, \rev{which in turn dominates the occurring integrals}. For $ \omega \to 0$ this leads to 

\begin{align}
\Sigma(\infty,z) &= \sum_{\7k} \hat{f}(\7k) \mathcal{V}(\7k,\infty,z)\;,\\
\mathcal{V}(\7k,\infty,z) &\approx \rev{\frac{1}{n}}\frac{ \hat{f}(\7k)+\sum_{\7q}  \hat{f}(\7k-\7q)  \mathcal{V}(\7q,\infty,z)}{G_0^{-1}(\7k,z)-\sum_{\7q} \frac{V^2(\7q,\7k)}{z-\epsilon(q)}+i \omega^3 \rev{\frac{ B_{\rm R}}{c_T^2} }h(\7k) }\;,
\end{align}
with $h(\7k) = \sum_{\7q} \7q^2 \frac{V^2(\7q,\7k)}{\epsilon(\7q)^2}.$ Considering the imaginary part of the self energy one obtains a term  linear in $\omega $. It originates from the sound pole in the denominator. Additionally, one gets a second term proportional to $\omega^3 $.  This term has a prefactor composed of a coefficient linear in \rev{$B_{\rm R}$} plus an  off-set, which is independent of damping and  results from the sound pole.  While the first term leads to the Debye spectrum, the second term 
leads to the $g_{\rm loc}= A_4 \omega^4$ vDOS of the quasi-localised modes for small frequencies.  \\
One can write down the coefficient $A_4$ of the vDOS of QLMs. But first, we need to calculate $\Im_{\omega^3}\Sigma$, which is the part of the imaginary part of the self energy that is proportional to $\omega^3 B_{\rm R}\rev{/c_T^2}$ 
We truncate the resulting series for  $\Im_{\omega^3}\Sigma$ after the first two contributions, but one can easily include more terms. One gets for $z \to 0$ \rev{
\begin{align}
\Im_{\omega^3}\Sigma=  \sum_{\7k} \frac{  \hat{f} (\7k)^2h(\7k) \rev{/n} }{	\left(\hat{f}(\70)-\hat{f}(\7k)-\sum_{\7q} \frac{V^2(\7q,\7k)}{z-\epsilon(\7q)}\right)^2}+  \sum_{\7k,\7q} \frac{2 \hat{f}(\7k)h(\7k)\rev{/n^2}}{	\left(\hat{f}(\70)-\hat{f}(\7k)- \sum_{\7q} \frac{V^2(\7q,\7k)}{\epsilon(\7q)}\right)^2} \frac{\hat{f}(\7k-\7q)\hat{f}(\7q)}{\hat{f}(\70)-\hat{f}(\7q)-  \sum_{\7l} \frac{V^2(\7l,\7q)}{z-\epsilon(\7l)}}  +\cdot \cdot \cdot \;,
\end{align}}
Here,  the factor $2$ is of combinatorial nature and results  from the choice of which denominator one takes the imaginary part of. Note, that one can not approximate $\omega \approx 0$ in the equation above, since this would remove the sound pole, which contributes to the Debye spectrum. All in all, this gives \begin{align}	\label{Supp_Coefficient_A4}
A_4= \rev{n}\frac{2}{\pi} \frac{	\Im_{\omega^3}\Sigma}{\epsilon(\infty)^2}\;  \rev{ \frac{B_{\rm R}}{c_T^2}  }\;.\end{align}
\rev{The leading order goes with $\propto n^{-5.5}$. In the main text, we compare \eqref{Supp_Coefficient_A4} with the simulation finding $ \frac{c_T^4 \omega_{BP}^2 }{B_R}A_4 \approx 0.01/2$  and the HET-prediction $A_4= \frac{4}{2}  \frac{B_R}{  \pi \omega_{BP}^2c_T^4} \frac{\omega_{BP}^2}{\omega_D^2} $ \cite{Wang_Stable_glasses, Ikeda_Phonon_transport,Schrimacher2007}. The additional factor $1/2$ arises since we consider only one transversal model. We find $c_T^4 \omega_{BP}^2A_4 / B_R\approx 0.045$ for $n=0.5$ and thus reasonable   agreement with the computational prediction. To compare to the HET-prediction, we set $\frac{\omega_{BP}^2}{\omega_D^2}\approx \frac{1}{36}$ as it has been found in \cite{Wang_Stable_glasses}. Our predictions exceeds the HET-prediction by approximately a factor of $\approx 2.5$}

 \section{Comparison with the planar theory \label{app_sec_Planar_theory}}
 
\rev{The planar theory of the ERM model  presented in \cite{ciliberti2003brillouin, Grigera2001} is easily obtained from our self-consistent theory \eqref{Self_consitent_model} by neglecting all the non-planar contributions of the $B$-and $C$-blocks. This leads to the following expression for the self energy \cite{goetschy2013euclidean}
\begin{align}
    \Sigma_P(\7p,z) = \frac{1}{n}\sum_{\7k} V^2(\7k,\7p)G(\7k,z) = \sum_{\7k} \frac{V^2(\7k,\7p)}{z-n\epsilon_0(\7k)-\sum_{\7q} V^2(\7q,\7k)G(\7q,z)}  \;.
\end{align}
This straightforward re-summation can not capture all the salient features of disordered materials. 
As shown by Refs.~\cite{Grigera2001,Ganter_Schirmacher} and others, see section \ref{Supp_Prof_Rayleigh}, the planar theory predicts hydrodynamic damping instead of Rayleigh-damping. One can quickly show this by restating the argument from \cite{Grigera2001}: For $\7p \to 0$, the main contribution to the self-energy arises from $k \gg p$. This gives
\begin{align}
     \Im \Sigma_P(\7p,z) = \frac{1}{n}\sum_{\7k} V^2(\7k,\7p) \Im G(\7k,z) \approx \frac{1}{n} \Im G(\infty,z)   \sum_{\7k} V^2(\7k,\7p) = - \pi g_\lambda(\lambda)   \sum_{\7k} V^2(\7k,\7p) \underset{\lambda, \7p \to 0}{\longrightarrow} -\frac{B_H}{n} \sqrt{\lambda} p^2\;.
\end{align}
Via  $ \Im \Sigma_P(\7p,z) \underset{\lambda, \7p \to 0}{\longrightarrow} -\frac{B_H}{n} \sqrt{\lambda} p^2\ $, we find the hydrodynamic damping coefficient. 
 In figure \ref{fig:Damping},  we compare the sound attenuation of the planar theory to the second order perturbation theory; for larger wavevector both lie close. } \\
 
\rev{The absence of Rayleigh-damping also implies that the strength of the sound attenuation  contributes linearly to the Debye-term of the vDOS instead of to the $\omega^4$-term.  To see this,  as \cite{ciliberti2003brillouin}
\begin{align}
\label{eq:Starting_Grigera_DOS}
    \begin{split}
        \frac{1}{G_\infty(z)}= \frac{z-\epsilon_P(\infty,z)}{n}-a G_\infty(z)-\frac{1}{n} \sum_{\7k} f^2(\7k) \Im G(\7k,z)\;,
    \end{split}
\end{align}
with $a=\sum_{\7k}/n $ as before, and $\epsilon_P(\7p,z)=n ( \epsilon_0(\7p)- \Re \Sigma(\7p,z) ) $ being the renormalized dispersion relation of the planar theory. In the hydrodynamic limit $(\omega \to 0)$, this  gives the quadratic equation
\begin{align}
    1= \Big[\frac{z-\epsilon_P(\infty)}{n}+i \omega B_H \sum_{\7k} k^2 \frac{f^2(\7k)}{\epsilon_p(\7k)^2} + i\frac{\omega f^2(\omega/c)}{4 \pi c^3} \Big]G_\infty(z)-  a G_\infty(z)^2 + \mathcal{O}(B_H^2)\;.
\end{align}
Here $\epsilon_P(\7p)=\epsilon_P(\7p,z=0)$ and  $c^2= \underset{\7p \to 0}{\text{lim}} \frac{\epsilon_P(\7p)}{p^2}$ being the speed of sound of the planar theory. Note, that in \cite{ciliberti2003brillouin} the authors approximated $G(\7k,z)  \approx G_0(\7k,z)$ in equation \eqref{eq:Starting_Grigera_DOS}, thus the term linear in $B_H$ is absent. For $n \to \infty$, one re-obtains their expressions since  $ \epsilon_P \propto n$ holds. For the density being sufficiently large, one gets  
\begin{align}
    G_\infty(\lambda) \approx \bigg( \frac{\lambda-\epsilon_P(\infty)}{n} +  i \omega B_H \sum_{\7k} k^2 \frac{f^2(\7k)}{\epsilon_p(\7k)^2} + i\frac{\omega f^2(\omega/c)}{4 \pi c^3} \bigg)^{-1}
\end{align}
It is noteworthy, that this expressions agrees with the one we would obtain, if we applied the low frequency approximation  as in Sect.~\ref{Supp_vDOS} to the planar theory. One gets the Debye-level
\begin{align}
  n  \frac{g_D(\omega)}{\omega^2}= \frac{2 n^2 B_H }{\pi \epsilon(\infty)^2} \sum_{\7k} k^2 \frac{f^2(\7k)}{\epsilon_p(\7k)^2} + \frac{ n^2 \hat{f}^2(0) }{2 \pi^2 \epsilon(\infty)^2 c^3 }\label{s39}
\end{align}
Consequently, the Debye spectrum increases with an increasing sound attenuation coefficient. We interpret this as the absence of quasi-localised modes.}  \\

\begin{figure}
    \centering
    \includegraphics{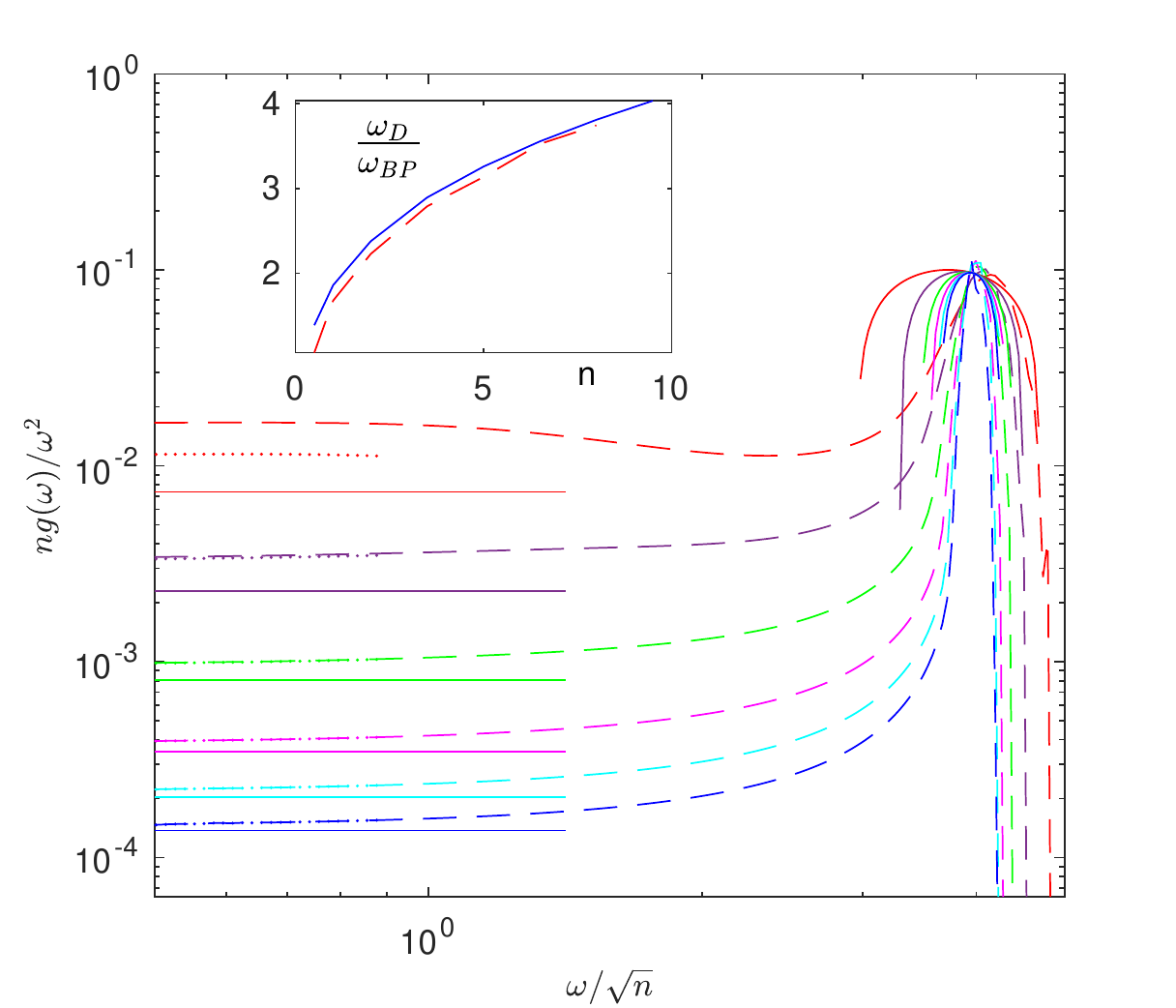}
    \caption{\rev{Comparison of the vDOS from Eq.~\eqref{Self_consitent_model}  (full lines) with the prediction of the planar theory (dashed lines) for different $n$ (see legend in Fig.~1). The dotted lines correspond to the Debye-level of the planar theory, Eq.~\eqref{s39}. The inset shows the associated ratio of Debye- over boson peak frequency. Here, the red dashed line shows the associated value of the planar theory. The blue solid line is the prediction of our model.}}
    \label{fig:Comparison_to_planar_theory}
\end{figure}

\rev{Lastly, in \cite{Grigera2001,ciliberti2003brillouin}, the boson peak was conceptualised as an excess over the Debye-vDOS which arises from an instability. Since the ERM-Model (Eq.~\ref{Def_Resolvent}) exhibits no such instability for a purely repulsive interaction, this instability is clearly an artifact of the approximations, as the authors of Ref.~\cite{Grigera2001,ciliberti2003brillouin} were aware.  Since recent simulations suggest \cite{Wang_Stable_glasses}, that the boson peak does not only occur in marginally stable systems, we identified the boson peak with the occurring semi-circle. We emphasise that this excess over the Debye-law is still disorder-induced, since the characteristic frequency  of the boson peak $\omega_{BP}$ is well below the Debye frequency $\omega_D$. 
For stable glass states, we compare the vDOS of the planar theory to our results in Fig.~\ref{fig:Comparison_to_planar_theory}. 
}

\section{References}
\rev{For the list of references see the main paper.}
\end{document}